\documentclass[preprint]{revtex4}

\usepackage{graphicx}

\begin{document}

\title{Electric Field Controlled Magnetic Anisotropy in a Single Molecule}

\author{Alexander S. Zyazin}
\email{a.zyazin@tudelft.nl}
\author{Johan W.G. van den Berg}
\author{Edgar A. Osorio}
\author{Herre S.J. van der Zant}
\email{h.s.j.vanderzant@tudelft.nl}
\affiliation
{Kavli Institute of Nanoscience, Delft University of Technology, PO Box 5046, 2600 GA Delft, The Netherlands}
\author{Nikolaos P. Konstantinidis}
\author{Martin Leijnse}
\author{Maarten R. Wegewijs}
\affiliation
{Institut f\"{u}r Theoretische Physik A, RWTH Aachen, 52056 Aachen, Germany, Institut f\"{u}r Festk\"{o}rper-Forschung, Forschungszentrum J\"{u}lich, 52425 J\"{u}lich, Germany, JARA- Fundamentals of Future Information Technology}
\author{Falk May}
\author{Walter Hofstetter}
\affiliation
{Institut f\"ur Theoretische Physik, Johann Wolfgang Goethe-Universit\"at, 60438 Frankfurt/Main, Germany}
\author{Chiara Danieli}
\author{Andrea Cornia}
\affiliation
{Department of Chemistry and INSTM, University of Modena and Reggio Emilia, via G. Campi 183, I-41100 Modena, Italy}

\begin{abstract}
We have measured quantum transport through an individual Fe$_4$ single-molecule magnet embedded in a three-terminal device geometry. The characteristic zero-field splittings of adjacent charge states and their magnetic field evolution are observed in inelastic tunneling spectroscopy. We demonstrate that the molecule retains its magnetic properties, and moreover, that the magnetic anisotropy is significantly enhanced by reversible electron addition / subtraction controlled with the gate voltage.   Single-molecule magnetism can thus be electrically controlled.
\end{abstract}

\maketitle

Rationally designed magnetic molecules~\cite{Kahn93,Gatteschi06} can be used as building blocks in future nanoelectronic devices for molecular spintronics~\cite{Bogani08}, classical~\cite{Affronte09} and quantum information processing~\cite{Leuenberger01,Loss07}. 
They usually have long spin coherence and spin relaxation times due to a weak spin-orbit and hyperfine~\cite{hyperfine} coupling to the environment.
Of crucial importance for applications is the ability to adjust the magnetic properties by external stimuli.
In bulk samples tuning magnetic properties by light has already been demonstrated~\cite{light}, but on the single-molecule level it has not been achieved. 
In addition, to tune magnetic properties on a single-molecule level, the use of local electric fields is preferred as it allows for a direct and fast spin-state control.

Addressing individual magnetic molecules on chip~\cite{Ralph08} has proven to be extremely challenging. 
Attempts to incorporate archetypal single-molecule magnets (SMMs) Mn$_{12}$ into a three-terminal device~\cite{Heersche06,Jo06} have been followed by observations that these complexes undergo electronic alterations when self-assembled on gold surfaces~\cite{Mannini08}. 
In this letter we demonstrate electric-field control over the magnetic properties of an individual Fe$_4$ molecule connected in a planar three-terminal junction. 
In the neutral state the bulk properties of this SMM are well documented~\cite{Accorsi06,Gregoli09} and more importantly they are retained upon deposition on gold ~\cite{Mannini09}.
From transport measurements we find that a Fe$_4$ molecule in a junction can still behave as a nanoscale magnet with the anisotropy barrier close to the bulk value.
In addition, upon reduction or oxidation induced by the gate voltage the barrier height increases; i.e., by charging, the molecule becomes a better magnet.

Characteristic of a SMM is its magnetic anisotropy, creating an energy barrier $U$ which opposes spin reversal; i.e., the high spin of the molecule points along a preferred easy axis, making it a nanoscale magnet. 
The anisotropy lifts the degeneracy of the ground high-spin multiplet, even in the absence of a magnetic field. 
The splitting of the lowest two levels is known as the zero-field splitting (ZFS); see Figure~1b. 
The magnetic anisotropy is described by the parameter $D$, which for the Fe$_4$ molecule in the bulk phase equals $D \cong 0.051-0.056$~meV; the ground state spin for this class of molecules $S=5$, the anisotropy barrier $U=DS^2 \cong 1.3-1.4$~meV and the ZFS is $(2S-1)D \cong 0.46-0.50$~meV~\cite{Accorsi06,Gregoli09}.

We use Fe$_4$ molecules with formula
[Fe$_4$L$_2$(dpm)$_6$]
(Hdpm = 2,2,6,6-tetramethyl-heptan-3,5-dione)
(see Figure~1).
Two derivatives, Fe$_4$Ph and Fe$_4$C$_9$SAc,
were synthesized by functionalizing the ligand H$_3$L = R-C(CH$_2$OH)$_3$
with R = phenyl and R = 9-(acetylsulfanyl)nonyl, respectively, and were prepared as described elsewhere~\cite{Accorsi06,Gregoli09}.
Their stability in a dry toluene solution at a 8 mM concentration was checked using spectroscopic techniques (see Supporting Information).
Nanometer-spaced electrodes were fabricated using self-breaking~\cite{ONeill07} of an electromigrated gold wire in a toluene solution of the molecules at room temperature~\cite{Osorio08}.
Three-terminal transport measurements were performed at a temperature $T = 1.6$~K.
By varying the gate voltage $V_{g}$ the molecular levels are shifted, thereby providing access to magnetic properties in adjacent charge states of the SMM.
Currently this is not possible with other techniques and therefore little is known about the magnetic properties of charged SMMs.

To quantify the anisotropy of an individual molecule we have performed transport-spectroscopy measurements in the Coulomb blockade regime. 
In total we have measured 648 devices, of which 48 showed Coulomb blockade signatures and two-dimensional conductance maps~\cite{Osorio08}  ($dI/dV$ versus gate and bias voltage) were measured on these.
The observation of molecule-related features (e.g. excitations) depends strongly on the dominant transport mechanism and electronic coupling to the leads. 
Typically, the strength of the electronic coupling in molecular junctions is significant so that the resolution of single-electron tunneling spectroscopy is limited by the tunnel-broadening.
In this case sharper higher-order inelastic cotunneling peaks~\cite{DeFranceschi01} may resolve the ZFS, but if their broadening due to Kondo correlations exceeds the ZFS, only a single, broad zero-bias Kondo peak remains, masking the low-bias ZFS features.
We identified 9 stable devices with indications of transport through a high-spin molecule: four devices showed broad Kondo peaks in two adjacent charge states, indicating that at least one of the charge states is a high-spin state. 
These Kondo features will be the subject of a subsequent paper.
Two samples showed a magnetic field-dependent transition at an energy scale of 1.5 meV and another one showed a low-bias current suppression indicative of spin blockade~\cite{Kramer95}, involving a high-spin ground state. 
The microscopic origins of these observations are not clear at the  moment; it should be noted, however, that the 1.5 meV value is close to the anisotropy barrier $U$.
Two devices showed transitions in the inelastic cotunneling spectra at an energy scale below 1 meV, close to the bulk value of the ZFS. 
In this paper we focus on these two devices.

In Figure 2a and 3a,b we present the measurements as differential conductance maps at zero magnetic field. 
Sample A (Figure~2) features the Fe$_{4}$Ph derivative and sample B (Figure~3) the Fe$_4$C$_9$SAc derivative.
The plots show conducting regions characteristic of single-electron tunneling (SET).
Outside the SET-regime, the molecule is in a well-defined charge state with $N$ electrons and the current is suppressed (Coulomb blockade).
Figure~2a and 3a,b clearly show a sizable conductance in the blockade regime due to tunnel processes of second (``cotunneling'') or higher order.
Here, inelastic cotunneling $dI/dV$ steps or peaks appear at a bias voltage $V_{b} = \pm \Delta / |e|$, and their observation allows for the determination of molecular excitation energies $\Delta$ in a particular charge state ~\cite{DeFranceschi01}. The charge state can be changed by the application of a gate voltage when crossing a highly conductive SET region. By making the gate voltage more positive, the electron number increases and the molecule is reduced; by making it more negative, the electron number decreases and the molecule is oxidized. 
 
In Figure~2a (sample A) three inelastic cotunneling lines are visible in the left charge state.
For large negative gate voltages these occur at $V \cong \pm0.6, \pm4.6$ and $\pm6.7$~mV.
For the right charge state there are lines at $V_{b} \cong \pm0.9$~mV and $\pm5$~mV.
For sample B cotunneling excitations appear at $V \cong \pm 0.9$~mV and $\pm5$~mV (Figure~3a) in the left charge state and at $V \cong \pm0.6~$mV and $\pm7$~mV (Figure~3b) in the right one.

Additional information on the nature of the low-energy excitations can be obtained by measuring their evolution in a magnetic field.
In Figure~2b,c and Figure~3c,d $dI/dV$ is plotted as a function of $V$ for different magnetic field values at
two adjacent charge states for each sample.
The energy of the lowest excitation increases with magnetic field without splitting and is symmetric upon field reversal (see Supporting Information).
Such behavior is a hallmark of ZFS described by the spin Hamiltonian~\cite{Gatteschi06,Sessoli93,Hirjibehedin07,Tsukahara09}
\begin{equation}
  \hat{H} = -D \hat{S}_z^2 + g\mu_B \vec{B} \cdot \hat{\vec{S}},
  \label{eq:ham}
\end{equation}
where the first term is the uniaxial magnetic anisotropy with the easy axis $z$. 
The second term is the Zeeman interaction of spin $\hat{\vec{S}}$ with magnetic field $\vec{B}$, where $g$ is the Land\'e factor and $\mu_B$ the Bohr magneton.
Importantly, model (1) predicts a non-linear dependence of the excitation energy on the magnetic field if the angle between the easy axis of the molecule and the magnetic field is substantial (see Supporting Information). This non-linearity is caused by the mixing of $\left|S_z\right\rangle$ states due to the presence of a transverse component of the field.

To quantitatively compare the data with the model, the excitation energy has been determined from individual $dI / dV$ curves.
For sample A (Figure~2b,c), we have taken the peak positions as the ZFS excitation energy and this energy is plotted versus magnetic field in Figure~2d,e.
For sample B, the inelastic cotunneling excitations have the form of conductance steps, reflecting a weaker molecule-electrode tunneling $\Gamma$ than for sample A.
We have fitted the measured $dI/dV$ to a Lambe-Jaklevic formula~\cite{Lambe68,Kogan04} and plotted the excitation energies in Figure~3e,f.

We first compare the data of sample A with the model. 
For the left charge state (Figure~2b) the energy of
the first excitation at $B=0$~T is close to the ZFS value (0.5~meV) in the bulk~\cite{Accorsi06}. 
We assume that this is the neutral state, and since the Fe$_4$ molecule maintains its ground state spin $S=5$ when deposited on gold~\cite{Mannini09}, we estimate $D \cong 0.06$~meV, so that $U \cong 1.5$~meV.
The increase of the excitation energy with $B$ is linear, implying an angle $\theta$ between $\vec{B}$ and the easy-axis below 45$^\circ$.
It should be noted that the angle cannot be controlled in our junctions and is expected to vary for different samples.
Using model (1) we have calculated the energy difference between the ground and the first excited state originating from the same side of the anisotropy barrier as a function of applied magnetic field. Using the angle $\theta$ as an additional parameter, the best fit of the data to the model, shown in Figure 2, yields g = 2.1 and $\theta = 0^\circ$.

Reduction and oxidation of the molecule inevitably change its magnetic properties, although it is not a priori clear in which way.
Three-terminal spectroscopic measurements can provide an answer since the ZFS and the change in the total spin upon charging can be obtained independently.
First, the difference in spin values of adjacent charge states can be determined from the shift of the degeneracy-point in a magnetic field (see Supporting Information).
We infer an increased $S=11/2$ in the reduced charge state of sample A.
From the measured ZFS we then find an enhanced $D \cong 0.09$~meV and $U \cong 2.7$~meV.
From the magnetic field dependence, we infer $g=1.8$ and $\theta=0^\circ$ (Figure 2e) when fitting the data to the model (1).
For sample B we observe the bulk ZFS value in the right charge state (Figure~3d,f) and therefore identify this as the neutral state with $S$ = 5 and $D \cong 0.06$~meV.
A clear non-linear Zeeman effect is now observed, indicating that the field is at a substantial angle
with the easy-axis.
We have fitted the data to the model obtaining $g=2.1$ and $\theta = 71^\circ$. 
Also for the left charge state the Zeeman effect is non-linear (Figure~3e) and a three-parameter fit 
yields $D = 0.09$~meV, $g = 2.0$ and $\theta = 60^\circ$.
Here, we have used $S=9/2$ for the oxidized state (see Supporting Information) and estimate $U \cong 1.8$~meV.
Note that for small magnetic fields the data show a deviation from the model, which is not yet understood.

In view of the rich SMM excitation spectrum the question arises
why only the ZFS excitation is observed up to a bias of several mV.
We have performed extensive calculations which indicate that this indeed should be the case (see also Supporting Information).
In Figure~4a,b we show $dI / dV$ maps calculated using quantum kinetic equations (KE), accounting for tunnel processes to first and second order in $\Gamma$~\cite{Leijnse08}. We model the low-energy spectrum of two successive charge states including the charge-dependent ZFS.
Figure~4 shows that for the experimental temperature the ZFS $dI / dV$-step is dominant,
showing a linear Zeeman effect for $\theta = 0^\circ$ (Figure~4a)
and a non-linear behavior for $\theta = 71^\circ$ (Figure~4b).
Numerical renormalization group (NRG)
calculations using a Kondo model~\cite{Romeike06a} cover the regime where the tunnel coupling dominates over the thermal energy
(in contrast to the KE approach).
Figure~4d,e show that the ZFS now appears as a peak due to the exchange scattering through the SMM indicating
significant Kondo correlations.

In summary, we have demonstrated electric-field control over the anisotropy of a single magnetic molecule in a three-terminal junction.
We found a stronger magnetic anisotropy upon both reduction and oxidation induced by the gate voltage.
This enhancement may be related to the alteration of single-ion anisotropy, which should be substantial when changing the redox state, but more studies including quantum chemistry calculations have to be performed to confirm this.
Our findings open the route to new approaches in tuning magnetism on a molecular scale and the possibility of manipulating individual magnetic molecules for future use in nanoelectronic applications.

This work was supported by NanoSci-ERA, FOM, DFG SPP-1243.

Supplementary information containing chemical characterization data, transport measurements and calculations details can be found at http://pubs.acs.org/doi/suppl/10.1021/nl1009603 free of charge.

\providecommand*{\mcitethebibliography}{\thebibliography}
\csname @ifundefined\endcsname{endmcitethebibliography}
{\let\endmcitethebibliography\endthebibliography}{}

\newpage
\begin{figure}[width=7in]
	\centering
		\includegraphics{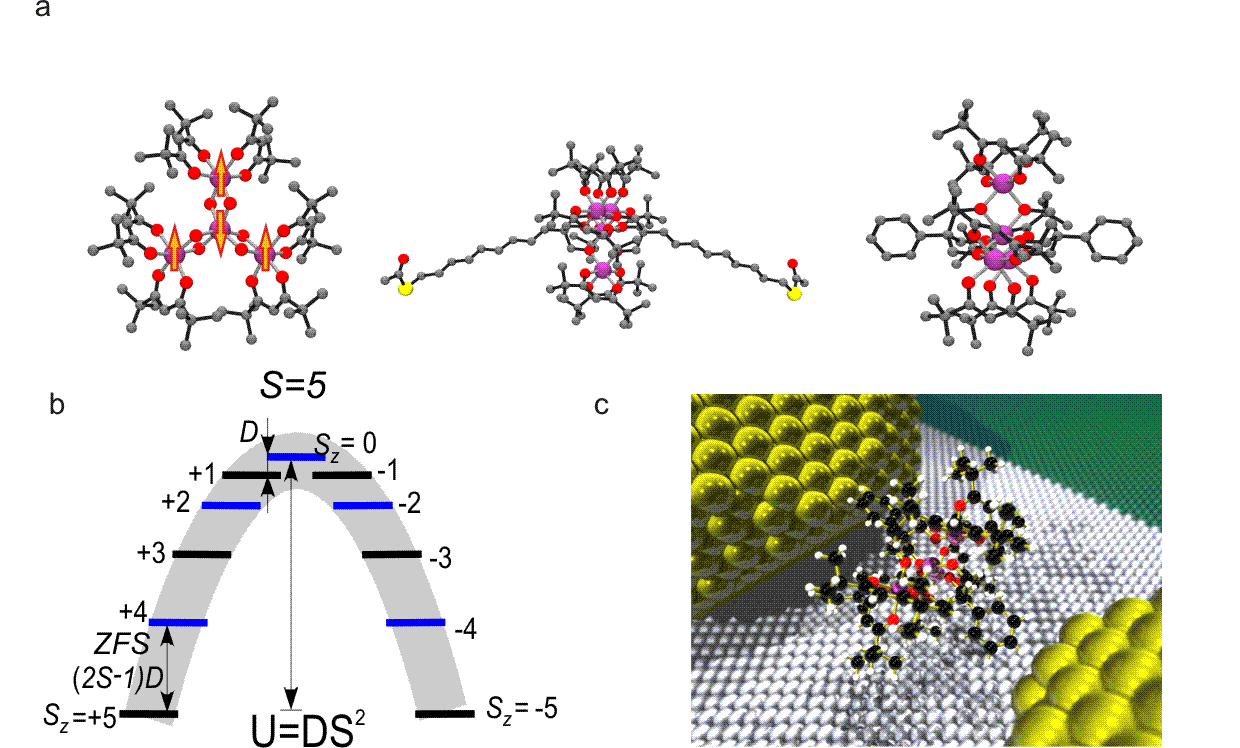}
		\caption{Fe$_4$ molecule. (a)
    Structure of the Fe$_4$ molecules (color code: iron=purple, oxygen=red, carbon=grey, sulfur=yellow).
    Left:
    Magnetic core with four $S=5/2$ iron (III) ions antiferromagnetically coupled to give a molecular spin $S = 5$.
    Center:
    Fe$_4$C$_9$SAc derivative.
    Right:
    Fe$_4$Ph derivative.
    (b) Energy diagram of the ground spin multiplet at zero field. The $S_z \neq 0$ levels corresponding to different orientations of the spin vector along the easy axis of the molecule are doubly-degenerate.
    The $S_z=+5$ and $S_z=-5$ states are separated by a parabolic anisotropy barrier $U$.
    An important property of the Fe$_4$ molecule is the large exchange gap
    to the next $S=4$ high-spin multiplet~\protect\cite{Carretta04} in the neutral state,
    which is 4.80 meV and 4.65 meV, for Fe$_4$Ph and Fe$_4$C$_9$SAc, respectively~\protect\cite{Accorsi06,Gregoli09}.
    Transport below a bias voltage of a few mV therefore only probes magnetic excitations
    of the ground high-spin multiplet, in contrast to the Mn$_{12}$ derivatives~\protect\cite{Heersche06}.
    (c) Drawing of a three-terminal junction with a single Fe$_4$Ph molecule bridging two gold electrodes (yellow)
    on top of an oxidized aluminum gate (grey).}
	\label{fig:vanderzant_fig1}
\end{figure}

\begin{figure}
	\centering
		\includegraphics{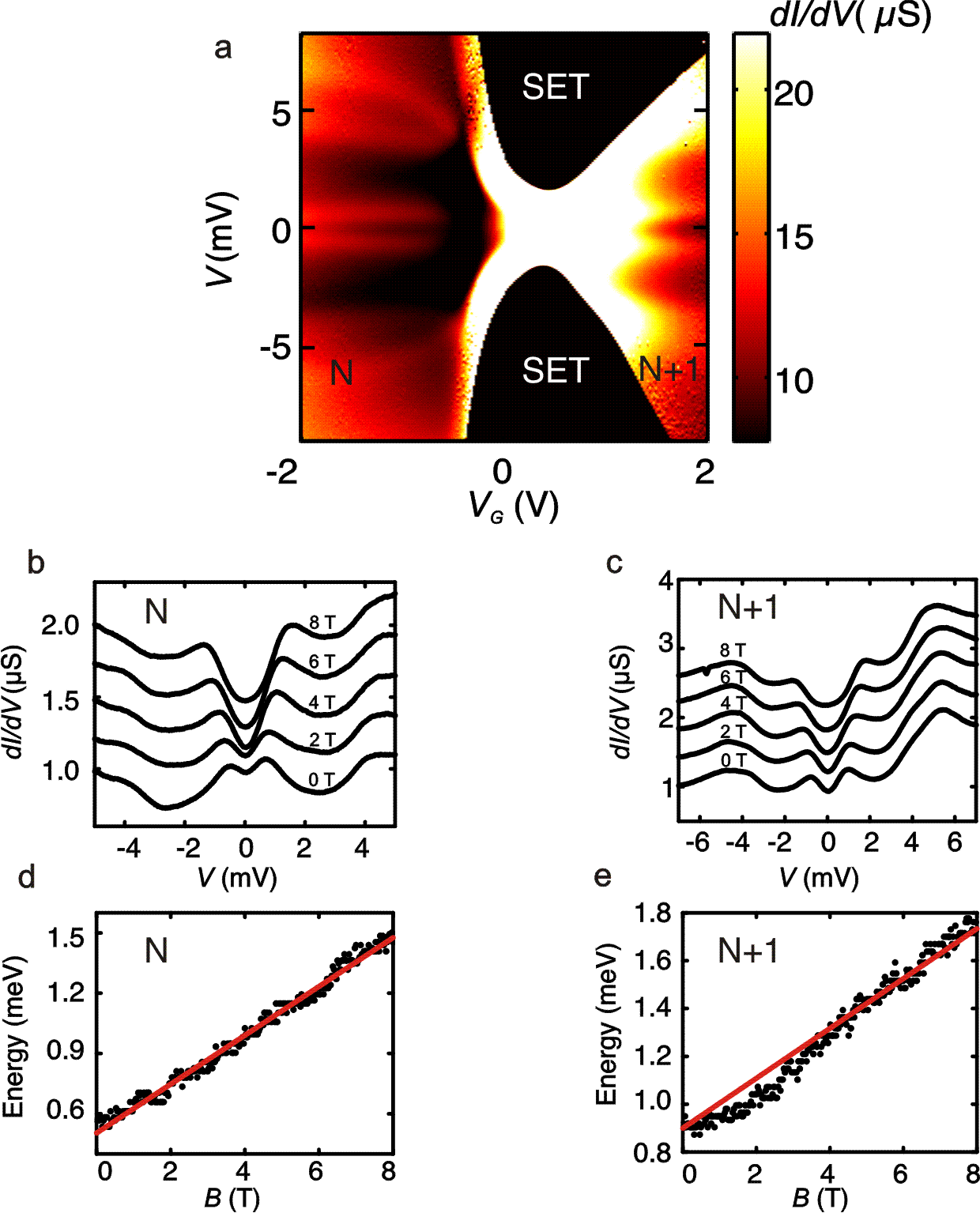}
	\caption{Characteristics of sample A. (a) Color plot of
    $dI / dV$ versus $V$ and $V_G$ at $T$~=~1.6~K and $B$~=~0~T.
    In the black regions in the middle of the plot single-electron transport (SET) is allowed.
    The lock-in amplifier saturates in these high conductance regions.
    (b)
    $dI / dV$ as a function of $V$ for $V_G$ = -1.5 V and various magnetic field values.
    Successive curves are offset by 250 nS.
    (c)
    Same as (b) for $V_G$ = 2 V and an offset of 400 nS.
    (d) and (e)
    Excitation energy as a function of magnetic field for the same $V_G$ values as in (b) and (c).
    Red lines are fits
    with
    $D$ = 0.06~meV, $g$ = 2.1, $S$ = 5 in (d) and
    $D$ = 0.09~meV, $g$ = 1.8, $S$ = 11/2 in (e) and $\theta$ = 0$^\circ$ for both cases.}
	\label{fig:vanderzant_fig2}
\end{figure}

\begin{figure}[width=7in]
	\centering
		\includegraphics{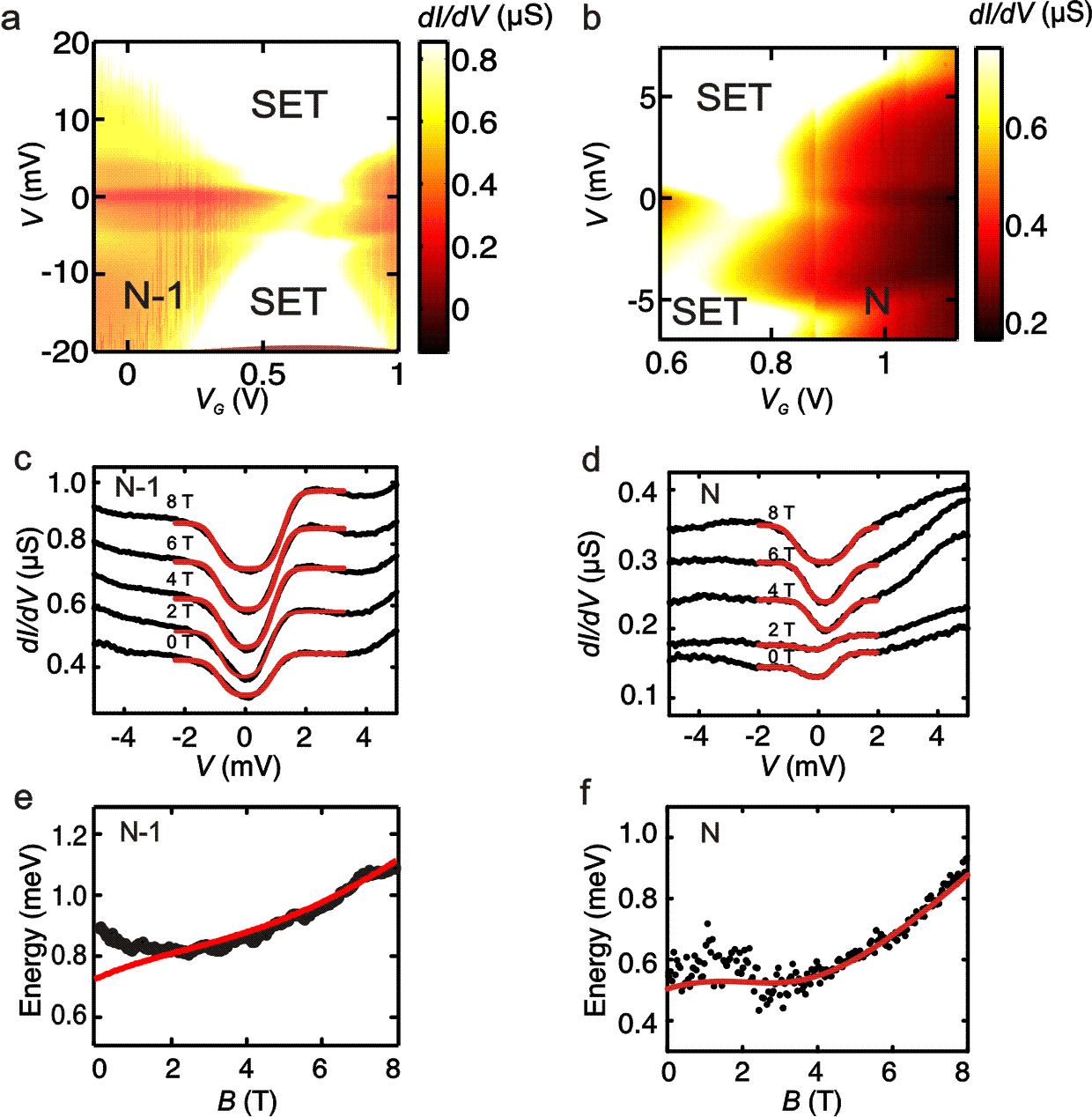}
		\caption{Characteristics of sample B. (a) and (b) Color plots of
    $dI / dV$ versus $V$ and $V_G$ at $T$ = 1.6~K and $B$ = 0~T
    highlighting the behavior in two adjacent charge states.
    (c)
    $dI / dV$ as a function of $V$ at $V_G$ = 0 V
    for successive magnetic fields $B$. Curves are offset by 125 nS.
    Red lines are fits to a Lambe-Jaklevic formula (see Supporting Information).
    (d)
    Same as (c) for $V_G$ = 1.6 V with curve offsets 50, 80, 140 and 220 nS.
    (e)
    Energy of the first excitation as a function of $B$ at $V_G$ = 0.1 V.
    The red line is a fit to model (1) for $S = 9/2$, $D = 0.09$~meV, $g = 2.0$ and $\theta = 60^\circ$.
    (f)
    Same as (e) for $V_G$ = 1.1 V.
    The red line is a fit for $S = 5$, $D = 0.06$~meV, $g = 2.1$ and $\theta$ = 71$^\circ$.}
	\label{fig:vanderzant_fig3}
\end{figure}

\begin{figure}
	\centering
		\includegraphics{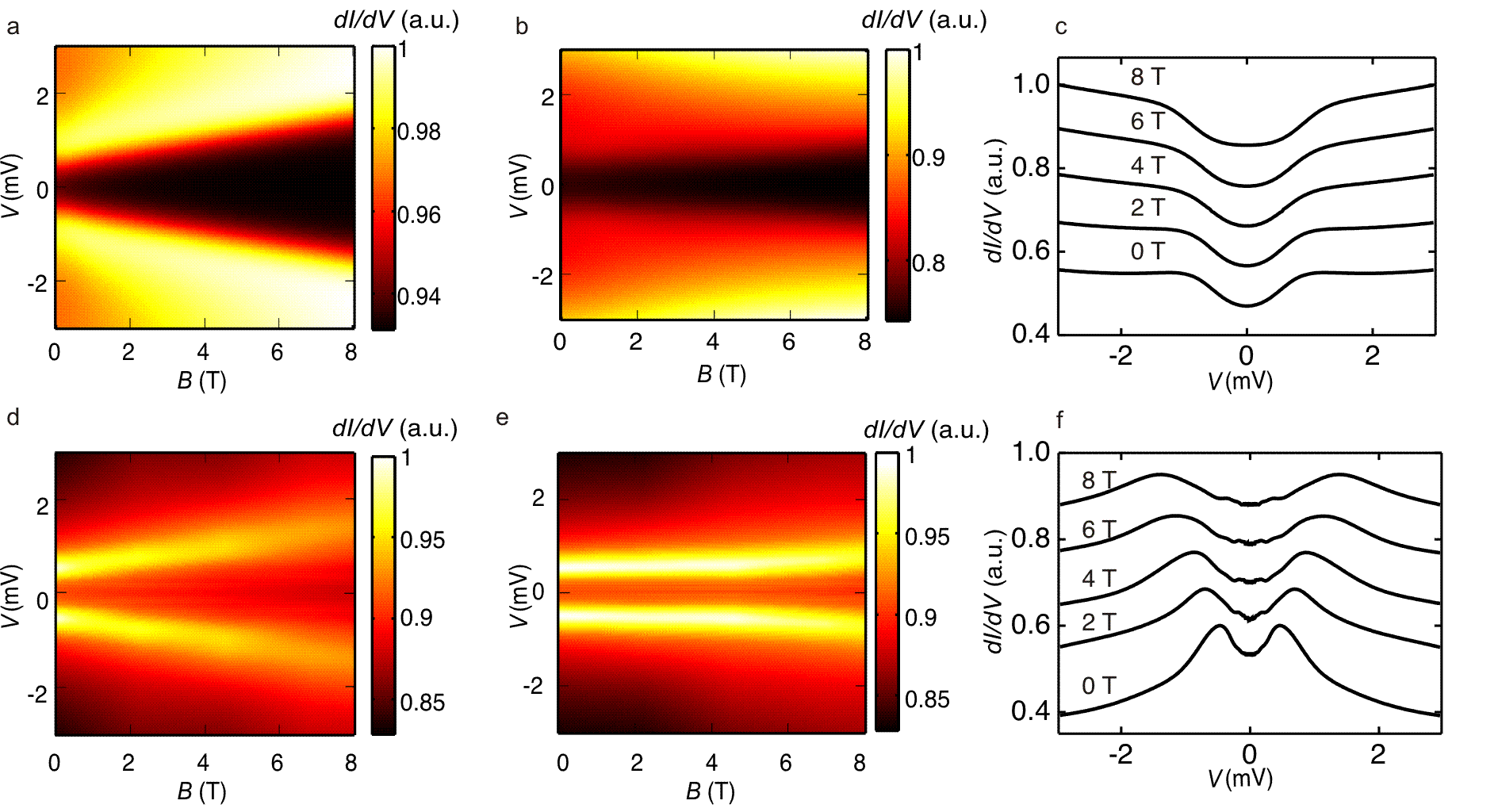}
		\caption{Calculated $dI/dV$ as a function of bias voltage and magnetuc field. (a) KE result for $\theta$=0$^\circ$ using the parameters estimated for sample A
    with electron temperature $T$~=~1.6~K and the gold conduction bandwidth $W$~=~8.1~eV.
    Shown is the field evolution of the inelastic cotunneling step for the $S=5$ state at gate voltage $V_G$ = -1.5~V.
    The conductance is scaled to its maximum value. 
    (b)
    Same as (a) but using the parameters estimated for sample B
    with gate voltage $V_G=1.6$~V and $\theta=71^{\circ}$.
    (c)
    $dI/dV$ traces taken from (b) corresponding to sample B.
    (d)
    NRG result for $\theta$=0$^\circ$ for the Kondo model using the same parameters as in (a)
    with zero temperature and an exchange tunneling constant $J = 0.1/\rho$ ($\rho$ is density of states).
    (e)
    Same as (d), but now for $\theta$ = 71$^{\circ}$.
    (f)
    $dI/dV$ traces taken from (d) corresponding to sample A.}
	\label{fig:vanderzant_fig4}
\end{figure}

\end{document}